\documentclass[10pt, conference, compsocconf]{IEEEtran}
\usepackage{cite}

\ifCLASSINFOpdf
   \usepackage[pdftex]{graphicx}
\else
\fi
\usepackage{amssymb}
\usepackage{pifont}
\newcommand{\cmark}{\ding{51}}%
\newcommand{\xmark}{\ding{55}}%

\begin{document}
%
\title{TwinCloud: Secure Cloud Sharing Without Explicit Key Management}




%
\author{\IEEEauthorblockN{
Kemal Bicakci\IEEEauthorrefmark{1},
Davut Deniz Yavuz\IEEEauthorrefmark{2},
Sezin Gurkan\IEEEauthorrefmark{3}}
\IEEEauthorblockA{Department of Computer Engineering\\
TOBB University of Economics and Technology}
\IEEEauthorblockA{\IEEEauthorrefmark{1}bicakci@etu.edu.tr, \IEEEauthorrefmark{2}st111101036@etu.edu.tr, \IEEEauthorrefmark{3}st111101054@etu.edu.tr}
}


\maketitle

\begin{abstract}
With the advent of cloud technologies, there is a growing number of easy-to-use services to store files and share them with other cloud users. By providing security features, cloud service providers try to encourage users to store personal files or corporate documents on their servers. However, their server-side encryption solutions are not satisfactory when the server itself is not trusted. Although, there are several client-side solutions to provide security for cloud sharing, they are not used extensively because of usability issues in key management.

In this paper, we propose TwinCloud which is an innovative solution with the goal of providing a secure system to users without compromising the usability of cloud sharing. TwinCloud achieves this by bringing a novel solution to the complex key exchange problem and by providing a simple and practical approach to store and share files by hiding all the cryptographic and key-distribution operations from users. Serving as a gateway, TwinCloud uses two or more cloud providers to store the encryption keys and encrypted files in separate clouds which ease the secure sharing without a need for trust to either of the cloud service providers with the assumption that they do not collude with each other. We implemented TwinCloud as a lightweight application and make it available as open-source. The results of our usability study show the prospect of the secure sharing solution of TwinCloud.

\end{abstract}


%
\IEEEpeerreviewmaketitle

\section{Introduction}
Cloud storage is a data storage model in which the data is stored in multiple remote servers and locations owned by a hosting company like Dropbox, iCloud or Google Drive. These cloud providers must keep the data accessible as well as confidential from other users. They offer services like easy file storage, sharing, syncing and collaboration among cloud users. Because of these useful services, today, both regular consumers and business organizations are widely using the cloud storage.

However, while cloud storage services provide these useful services, more and more personal and confidential data is being exposed to privacy and security vulnerabilities.

iCloud, for instance, was hacked by attackers using phishing and brute-force tactics leading to the subsequent posting of private celebrities' photos online \cite{1}. Also, in June 2011, several Dropbox cloud users were able to access personal files of other users without authorization due to a bug in the authentication mechanism \cite{2}.

Cloud providers have developed new security methods to protect user's data from attackers. For example, Dropbox and iCloud started to use two-step verification where the idea is to combine "something you know" (password) with "something you have" (phone) to add an extra layer of security \cite{3}, and also Dropbox uses Secure Sockets Layer (SSL)/Transport Layer Security (TLS) to protect data in transit between client apps and Dropbox servers \cite{4}. Moreover, Google Drive stores the data randomly distributed across multiple machines which provide another security level \cite{6}.

However, these security mechanisms are not sufficient and only protect users from third party unauthorized access, not from the cloud storage provider's access to confidential files. Google states in its Terms of Service that when you upload or submit a file, you give Google a license to use, modify, publish and distribute the content of the file \cite{5}. Furthermore, Google analyzes the file content and uses for advertising. Dropbox has also a similar term of service for the files that you put in their storage. Moreover, according to Snowden's documents \cite{19}, NSA has a secret program called Prism which the agency collects sensitive data from Google, Facebook, Apple, Yahoo and other US Internet giants.

In order to protect the customer's data stored in the cloud, two models are used among cloud service providers and third party cloud applications:

\begin{enumerate}
  \item Server-side encryption: Data is encrypted after uploading to the cloud.
  \item Client-side encryption: Data is encrypted before uploading to the cloud.
\end{enumerate}

For server-side encryption, Dropbox uses 256-bit Advanced Encryption Standard (AES) to encrypt uploaded files before writing to the disk \cite{4}. Google Drive also uses 256-bit AES, and each encryption key is itself encrypted with a regularly rotated set of master keys \cite{7}. These security measures protect the data from attackers, but cloud providers have the encryption key and the files get decrypted on their servers' every time they are accessed. Moreover, their administrators can see user's files, and so can anyone who manages to gain access to their systems. For instance, in 2010, a Google employee accessed several Gmail and Hangouts accounts to spy and harass people \cite{8}. Thus, server-side encryption is not sufficient itself for security and need to be supported by other security measures like client side encryption.

Client-side encryption will eliminate the above problem since files are encrypted before uploading to remote cloud servers with an encryption key that only a specific user knows. There are several commercial cloud encryption programs such as nCrypted, Sookasa, Tresorit and BoxCryptor which provide client-side encryption. However, this method reduces the usability of cloud services. With client-side encryption, users could not share their files as easily as before because they also need to share the encryption key with other users from a secure channel. In order to solve the key distribution problem, several PKI-based methods were proposed \cite{10}, \cite{11}, \cite{13}, \cite{15}. However, PKI-based solutions have some drawbacks. They are costly because they need to get a certificate from a CA. On the overall, it is well-documented that managing cryptographic keys is not manageable for average computer users \cite{26}.

There are several methods used by cloud providers to share files. File owners can use one of the following sharing methods to share their files with business partners, friends and coworkers \cite{9}:

\begin{enumerate}
  \item Public sharing: File is shared with a public URL that everyone can access.
  \item Secret-URL sharing: File is shared by sending a private sharing URL to specific users.
  \item Private sharing: File owner must specify who can access the shared file and cloud service provider authenticates the specified users while accessing the file.
\end{enumerate}

Both Dropbox and Google Drive supports all of these sharing methods (however as of this writing private and public sharing is only applicable for folders, not for files on Dropbox).

The sharing methods listed above are used only for sharing the original file and are not yet used for key sharing. Thus, after encryption, using cloud providers' services, it is only possible to share the encrypted file. Encryption key needs to be shared from another channel. Finding a secure channel to share the key is a problem with serious usability challenges.

In this paper, we present a novel solution, TwinCloud, that uses client-side encryption and private sharing method. Cloud users could continue to use the cloud services as before in a simple way. Users are able to share their files securely using the TwinCloud application which maintains the encrypted file and the encryption key in separate clouds where users can do all file operations from this single application without a need for explicit key management.

TwinCloud generates a key and encrypts the file with this key before storing in cloud servers. Then, it uploads the encryption key to a cloud and encrypted file to another cloud. Users easily share a file to a specific user by using the TwinCloud's user interface while TwinCloud, in the back-end, shares the encryption key and the encrypted file to the specific user using cloud providers' file sharing features.

The rest of this paper is organized as follows: Section II discusses the related work. Section III describes our solution and the properties of TwinCloud. Section IV provides a comparison of our solution to other solutions. Section V presents the results of our usability study of TwinCloud. Section VI discusses several extensions and provides ideas for future work. And finally, Section VII concludes the paper.


\section{Related Work}
This section examines preceding academic and industrial work related to secure file sharing on clouds.
Zhao et al. \cite{12} proposed a trusted data sharing method over untrusted cloud storage providers. This method can be described as follows:

\begin{enumerate}
\item Alice encrypts her data and keeps the data on a cloud storage.
\item Bob sends a request to Alice asking for access permission to the data.
\item Alice sends a credential to the cloud storage provider for the re-encryption of the data and sends a credential for Bob to decrypt the re-encrypted data with his private key.
\item When Bob acquires the re-encrypted data from the cloud storage provider, he decrypts it.
\end{enumerate}

The credentials are sent using Public Key Cryptography (PKC).

Jivanyan et.al \cite{20} proposed a proxy re-encryption scheme for secure file sharing without certificates. The solution uses PKC but does not use a PKI. In this solution, Alice shares the encrypted file with Bob as follows:

\begin{enumerate}
\item Alice generates a proxy key for Bob and gives the key to the proxy server.
\item Alice creates a permission object to indicate the file is allowed to be viewed by Bob.
\item Bob extracts the encrypted key file from the beginning of the file. The file key is encrypted via Alice's public key.
\item Bob sends the encrypted key file to the proxy server and asks to be re-encrypted.
\item Proxy server checks whether Bob was given permission for the file. If he has the permission, the proxy server re-encrypt the key file with proxy-key given by Alice. The new encrypted key file sent back to Bob.
\item Bob decrypts the re-encrypted key file with his private key. The content of the file can be decrypted with the decrypted key file.
\end{enumerate}
When Alice needs to share her files with many users, she just creates corresponding
proxy-keys and the permission objects indicating which user can access
the given file. There is no need to encrypt the file encryption key with all users
public keys which is one of the advantages of this system.

Yin et.al \cite{11} proposed another efficient secure data storage scheme using Elliptic Curve Cryptography (ECC) based PKI. Before using this solution, users need to authenticate to a Certificate Authority (CA) and register for a certificate. When users want to share a file with other authenticated users, the user first encrypts the data with his symmetric key and encrypts the symmetric key with the private key of his/her registered certificate. After the encryption process, authenticated users can obtain the public key and decrypt the encrypted key to get the symmetric key. The symmetric key, then, will be used to decrypt the file. The main advantage of this solution is the efficiency of preferring ECC over other algorithms like RSA. Because of the small key length, this solution requires less computation and communication cost.

Boxcryptor \cite{13} is a widely used commercial tool to protect files in the cloud and it supports many cloud service providers such as Dropbox, Google Drive, Microsoft OneDrive and Box. It uses AES with a key length of 256 bits, CBC (Cipher Block Chaining), PKCS7 padding and RSA with a key length of 4096 bits and OAEP padding. Boxcryptor uses following steps to share file between Alice and Bob \cite{14}:

\begin{enumerate}
\item Alice requests Bob's public key from the Boxcryptor Key Server.
\item Alice encrypts the file key with Bob's public key.
\item Alice writes the new encrypted key file.
\item The cloud storage provider syncs the modified encrypted key file.
\item Bob uses his private key to decrypt the file key.
\item Bob uses the file key to decrypt the file.
\end{enumerate}

Boxcryptor also uses a trusted CA.

Tresorit \cite{15} is a cloud provider that uses client-side encryption. They use PKI and PKC to share the encryption key. Tresorit does not fully trust any other certificate issuer, thus certificates issued by Tresorit company itself with the Tresorit User CA certificate. Users can share files by creating Tresor (a shareable secure online folder) that contains the files they want to share or by creating and sending encrypted links to other users.

There are several other sharing methods used with client-side encryption and nearly all of these methods \cite{10}, \cite{21}, \cite{22} use PKC or PKI-based solutions. However, there are some disadvantages of these solutions in terms of usability and security. Managing a trusted PKI is costly and hard to maintain. It also burdens the users with handling cryptographic operations. Moreover, if an attacker manages to obtain the private key of a CA certificate, the attacker can see all the files and its contents. In above cases, the attacker could potentially be someone from the cloud service providers. The fact that the certificates were issued by cloud providers verifies that they also can view user data whenever the keys are used for encryption/decryption purposes \cite{16}.

By realizing this, we design TwinCloud that does not require a trusted third party like a CA and does not use a PKI structure or PKC. It uses only symmetric key cryptography. It also does not require users to handle key management.

\section{Solution}
\subsection{Solution in a Nutshell}
TwinCloud is a lightweight application that uses client-side encryption and private sharing (see section I).

Our solution requires at least two cloud service providers. These cloud providers must provide simple file operations such as uploading, downloading a file and private file sharing. We use one of the cloud service providers in order to store the encrypted file and the other one to store the encryption key. This ensures that both cloud providers cannot see the file content by themselves. We assume that two cloud providers do not collide to bring encrypted file and the key together to decrypt the file. Against such a threat, the solution could be extended to using more than two clouds. Alternatively, encryption keys could be stored in a private, internal cloud. Companies can use their enterprise clouds (which does not need to have a high storage capacity) to store keys and a cloud service provider to store encrypted files. 

For file sharing, the application shares the encrypted file and the encryption key to the specified specific user simply by using cloud providers' file sharing features. With this way, only the specific user can access the encrypted file and the encryption key, and can use them to decrypt the shared file. The solution does not require a complex PKC based key exchange protocol. 

Authentication of users is managed by cloud service providers which eliminate the need for a PKI structure or an additional server to store user information. However, a custom designed authentication method of TwinCloud can be applied.

Our solution does not store user accounts and related key paths to encrypted files on the client side. User accounts are managed by cloud service providers. Key files and encrypted files are uploaded with names that can be matched in order to reach them easily after uploading.

TwinCloud uses symmetric key cryptography for file encryption. The encrypted files are stored in the cloud storage. It deletes the key files and the encrypted files from the local computer's temporary folder after uploading and downloading operations are completed.

\begin{figure}[!t]
\centering
\includegraphics[width=3.2in]{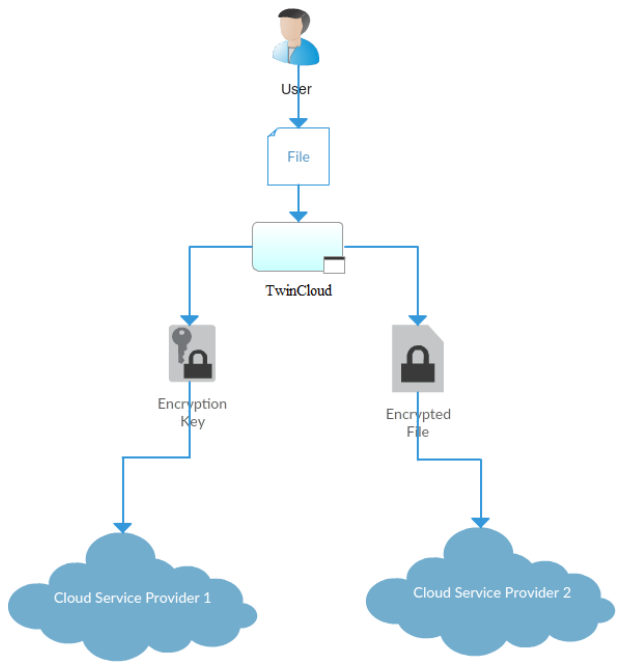}
\caption{Uploading a file to cloud using TwinCloud.}
\label{fig_1}
\end{figure}

The basic explanation of uploading a file is illustrated in Fig. \ref{fig_1}.

In summary, TwinCloud provides a single user account and a uniform user experience to execute all file operations. In the backend, the application manages two different accounts from two separate cloud service providers. While users are uploading their files through the application, files are not stored by our application, rather stored in external cloud drives.

\subsection{Solution Details}
In our solution, we choose two cloud service provider; Dropbox and Google Drive as they are preferred the most among cloud users. Moreover, their application program interfaces' (APIs) are easy to understand and highly functional.

TwinCloud uses Google Drive to store the encrypted file and Dropbox to store the encryption key. We use Java programming language for implementation. Dropbox API and Google Drive API are used for cloud file operations. Our application uses 256 bit AES with CBC and PKCS7 padding. Encryption keys are randomly generated by Java Key Generator utility. For authentication and authorization, we use OAuth2.0 \cite{17} protocol. Sign up and login operations to Google Drive and Dropbox are also controlled by the application.

There are seven main operations implemented in our application; sign up, login, upload a file, download a file, delete a file, share a file and unshare a file.

\textbf{Sign up:} TwinCloud uses a user interface similar to Google's Sign Up interface for Sign up. When a user enters a username and password, application derives two passwords from the original password. This derivation is achieved by calculating two different hashes. We use Hash(Password || Username || URL) function to derive the passwords. When generating the Dropbox password, the URL field is concatenated with Dropbox URL and Google URL is used to generate the Google Drive password. The resulting hashes, then, processed to fit in the provider's password fields. TwinCloud uses the username and the first password to create a Google Account and uses the second password to create a Dropbox account with the same username. Thus, if an attacker obtains one of the passwords, since the other password cannot be obtained without knowing the original password, it will not be possible to authenticate in both clouds and decrypt the file. Furthermore, this prevents a plausible attack when an authentication bug occurs, as occurred before in Dropbox \cite{2}, in one of the cloud service providers. Hence, we think that this adds another level of security to our cloud solution.

Selenium, a tool for automating web applications \cite{18}, is used for creating Google and Dropbox accounts.

\textbf{Login:} Users login to the system using their usernames and passwords. TwinCloud again derives the Dropbox and Google passwords to login to cloud service providers. After logging in, the user is asked whether they are willing to grant the permissions that the application is requesting. These permissions are required by Google Drive and Dropbox in order to do the necessary file operations. OAuth 2.0 authentication flow is used for obtaining the access token.

\begin{figure}[bhp]
\includegraphics[width=\columnwidth]{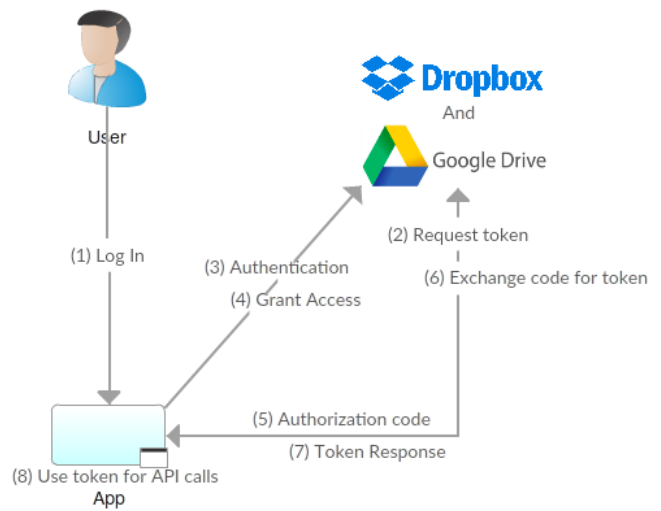}
\caption{Detailed login flow in TwinCloud.}
\label{fig_2}
\end{figure}

Fig. \ref{fig_2} shows the authentication flow. First, user logs in to the TwinCloud by entering his/her username and password (1). TwinCloud, in the backend, performs the login to Dropbox and Google Drive by following these steps: TwinCloud requests an access token from Google Drive and Dropbox separately (2). Dropbox and Google Drive request an authentication. This authentication is performed by the application (3) and the program automatically grants the access (4). Dropbox and Google Drive returns an authorization code (5) which will be used to get the access token. TwinCloud uses the authorization code to exchange the access tokens (6). Finally, Dropbox and Google Drive returns the access tokens (7). The application saves the access tokens in order not to repeat these steps in every login. Authentication and Grant Access operations are also implemented using Selenium.

Similar to other cloud applications such as Dropbox and Google, if a user has already an account on TwinCloud and opens the program for the first time, the application automatically starts to download the stored files into the local computer and decrypts them.

\textbf{Upload file:} Files can be uploaded to the cloud using simple drag-and-drop operations. TwinCloud executes the following steps while uploading a file named \textit{hello.txt}:
\begin{enumerate}
\item Generate the file encryption key (k).
\item Encrypt the file with key k.
\item Save the encryption key to a file named \textit{hello.txt.key} to the temporary folder.
\item Save the encrypted file as \textit{hello.txt} to the temporary folder.
\item Create a folder named \textit{hello.txt\_keyFolder} in Dropbox. (Since Dropbox does not support private sharing over files, application creates a folder with the same name and uploads the key file into the folder for future file sharing operations.)
\item Upload \textit{hello.txt.key} file to \textit{hello.txt\_keyFolder/hello.txt.key} in Dropbox.
\item Upload encrypted \textit{hello.txt} file to \textit{/hello.txt} in Google Drive.
\item Delete \textit{hello.txt.key} from the temp folder.
\item Delete encrypted \textit{hello.txt} file from the temp folder.
\end{enumerate}

\textbf{Download file:} Selected files can be downloaded using their filenames. TwinCloud executes the following steps while downloading a file named \textit{hello.txt}:
\begin{enumerate}
\item Download the key file from Dropbox with the \textit{hello.txt\_keyFolder/hello.txt.key} path to the temp folder.
\item Download the encrypted file from Google Drive to the temp folder.
\item Decrypt the encrypted file (hello.txt) with the key file (hello.txt.key) and save to the local computer.
\item Delete \textit{hello.txt.key} from the temp folder.
\item Delete encrypted \textit{hello.txt} file from the temp folder.
\end{enumerate}
\begin{figure}[bhp]
\includegraphics[width=\columnwidth]{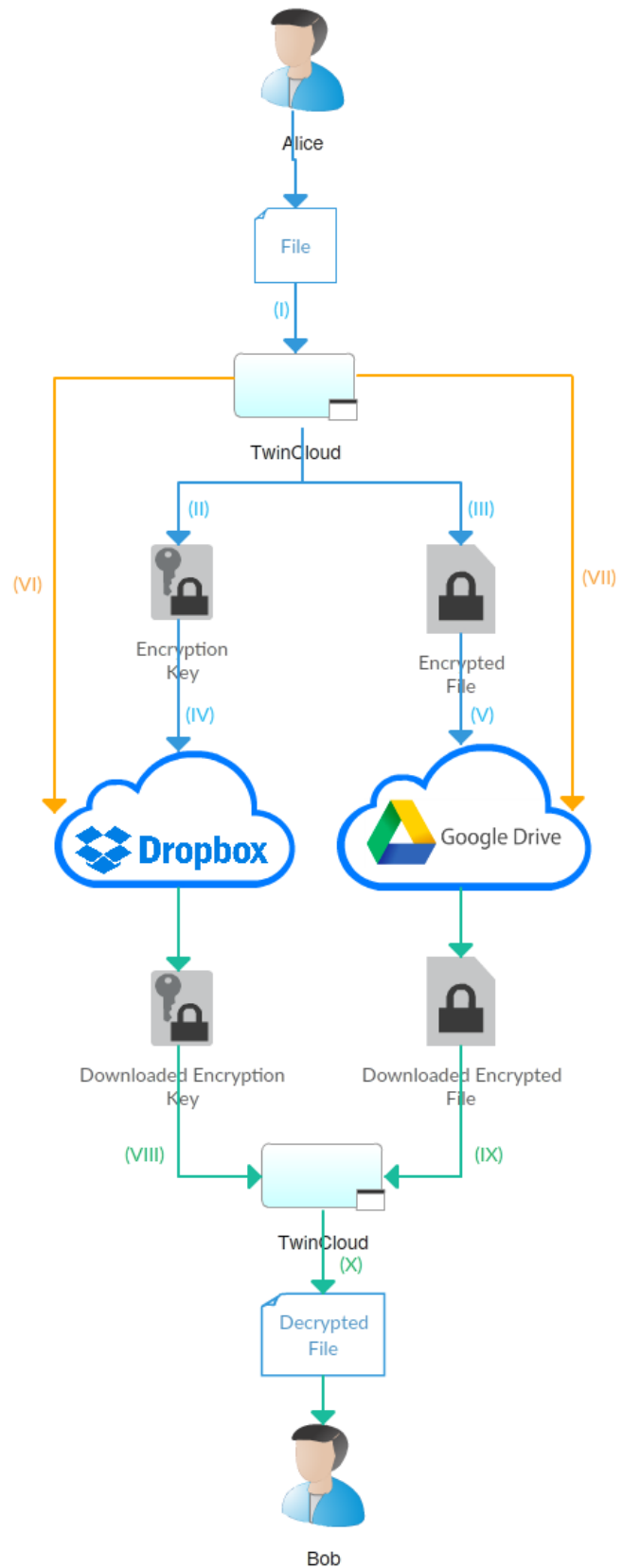}
\caption{Visualization of the upload, share and download operations in TwinCloud.}
\label{fig_3}
\end{figure}
\textbf{Delete file:} Selected files can be deleted from cloud storage using their filenames. TwinCloud executes the following steps while deleting a file named "hello.txt":
\begin{enumerate}
\item Delete the key folder from Dropbox with the \textit{/hello.txt\_keyFolder} path. The folder also deleted from cloud service provider's trash.
\item Delete the encrypted file \textit{hello.txt} from Google Drive. The file is also deleted from cloud service provider's trash.
\end{enumerate}
\textbf{Share file:} Selected files can be shared with specific users. TwinCloud executes the following steps for sharing a file named \textit{hello.txt} with the user Bob:
\begin{enumerate}
\item Share folder \textit{/hello.txt\_keyFolder/} in Dropbox with the user Bob using Dropbox API.
\item Share file \textit{hello.txt} in Google Drive with the user Bob using Google Drive API.
\end{enumerate}
After this operation, when Bob list its files, he will directly see the file \textit{hello.txt} in its file list because Google Drive API shows the shared files along the uploaded files when listing all files. Users can also select specific permissions while sharing files such as; can edit and can read. Moreover, files can be shared more than one person.

\textbf{Unshare file:} After sharing a file, the user can revoke this permission. TwinCloud executes the following steps while unsharing a file named \textit{hello.txt} from the user Bob:
\begin{enumerate}
\item Remove Bob's permission to access folder \textit{/hello.txt\_keyFolder/} in Dropbox using Dropbox API.
\item Remove Bob's permission to access file \textit{hello.txt} in Google Drive using Google Drive API.
\end{enumerate}

Fig. \ref{fig_3} shows the upload, share and download operations. 
After login, Alice uploads a file to the TwinCloud (I). The application generates a key (II) and encrypts the file (III). Then, these files are uploaded to Google Drive and Dropbox using APIs as seen in (IV) and (V). After successfully uploading the file, Alice shares the file with Bob using Dropbox API (VI) and Google Drive API (VII). This will allow Bob to see the shared file from its application. Bob requests to download the file from the application. Our application downloads the encryption key from Dropbox (VIII) and encrypted text from Google Drive (IX). After completing the downloads, the application decrypts the file (X). From now, Bob can also see the content of the shared file.

\begin{figure}[!t]
\centering
\includegraphics{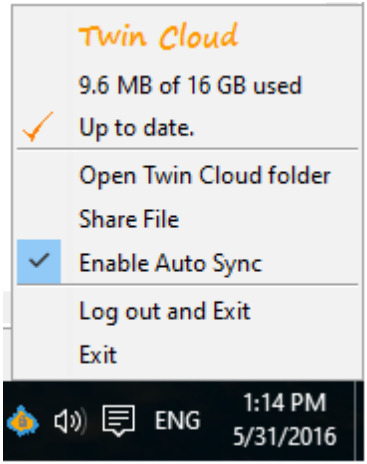}
\caption{TwinCloud start-bar menu.}
\label{fig_4}
\end{figure}

\begin{figure}[!t]
\centering
\includegraphics[width=\columnwidth]{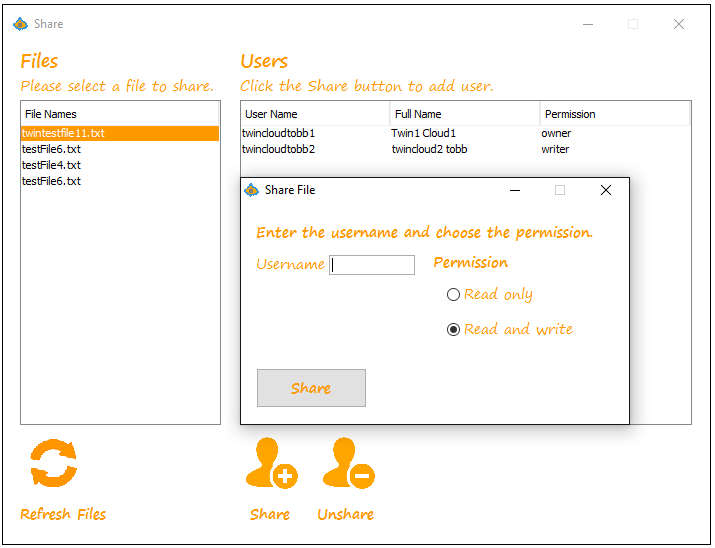}
\caption{TwinCloud's share window.}
\label{fig_5}
\end{figure}

We implemented TwinCloud in Java. Users can operate TwinCloud from a start-bar menu as shown in Fig. \ref{fig_4} and share their files from the application as shown in Fig. \ref{fig_5}. TwinCloud is an open source application and can be found on Github\cite{24}.
\section{Comparison}
Table. \ref{table_1} provides a comparison of TwinCloud and other solutions.

PKI-Based solutions use PKC and need a trusted third party e.g., CA. The key management is complex as discussed before. For easy key recovery and file recovery, this solution needs to store backups of these files whereas in our system it is ensured by the cloud providers. Private sharing and secret URL sharing (see Section I) could be supplied by PKI-based solutions. Also, the integrity of files is protected. Traditional Client-side Encryption uses only symmetric key encryption. Users, before uploading their files to the cloud, encrypt files using third-party encryption programs. Thus, this does not require a trust to the cloud service provider. This method requires users to manage their own keys for file operations and sharing. Therefore key management is again complex. Users need to store replicas of the encryption key and the encrypted file for file recovery. Private sharing and Secret-URL sharing are not directly supported because of the key exchange issues. Moreover, users need to use other programs to protect integrity.

Server-side encryption is provided by cloud providers. Files are encrypted while uploading to the cloud. Cloud providers can easily access to users'€™ files. It requires a full trust to the cloud service provider. Encryption keys and encrypted files are stored replicated for easy recovery. They support private sharing and secret-URL sharing. Integrity check is performed by the cloud provider.

\begin{table*}[t]
  \centering
  \begin{tabular}{*{15}{c}}
   & TwinCloud & PKI-Based & Traditional Client-Side Encryption & Server-Side Encryption \\
    \hline\\
    Requires a trust to the cloud service provider & \xmark & \xmark & \xmark & \cmark\\
    Requires a trusted 3rd party & \xmark & \cmark & \xmark & \xmark\\
    Public key cryptography & \xmark & \cmark & \xmark & \xmark\\
    Symmetric key cryptography & \cmark & \xmark & \xmark & \cmark\\
    Easy key management and key exchange & \cmark & \xmark & \xmark & \xmark\\
    Private sharing & \cmark & \cmark & \xmark & \cmark\\
    Secret-URL sharing & \cmark & \cmark & \xmark & \cmark\\
    Integrity check & \cmark & \cmark & \xmark & \cmark\\
  \end{tabular}
  \caption{Comparison of TwinCloud to other solutions. }
  \label{table_1}
\end{table*}

In PKI-Based systems security of the system depends on CA and cloud provider. In traditional client-side encryption, it depends on key sharing channel and encryption program. For server-side encryption, security requires a trust to cloud provider.

\section{Usability Of TwinCloud}
Through a user study, we evaluated the usability of the TwinCloud and compared it with Tresorit \cite{15}. We report the results in this section.

Tresorit is an end-to-end encrypted cloud storage. Users can share their files using encrypted links with an additional password protection option. They can also set expiration date and limit the number of downloads. Since it has the encrypted link option for sharing and a simple user interface we chose Tresorit to compare with TwinCloud.

Encrypted links contain fragment identifiers. When an encrypted link is opened, the web browser requests the encrypted data from Tresorit servers and sends the URL (without the fragment) to the server. As the fragment itself will not be sent, the server cannot access the data sent, but the browser, by combining the encrypted file and the key contained in the fragment, is able to decrypt it \cite{25}.

The security of the sharing mostly relies on the security of the channel where users send the encrypted link (in most cases this channel is e-mail). In a broad sense, TwinCloud and Tresorit offer the same level of security. While, the later assumes that Tresorit and email service provider do not collude, the former makes the same assumption for the two cloud service providers. 

E-mail servers of Google or Microsoft scan and parse the mail links to detect online scam and fraudulent links. However, this can lead to the leakage of information where government agencies can request to collect and access to users encrypted files \cite{25}. In case the URL is exposed, shared files lose their protection. Moreover, a malicious browser extension can expose the encrypted link. These kinds of attacks cannot be applied to our solution because we do not have an encrypted link option. The optional password protection feature will add more security to sharing if users choose to transmit the password from another channel and use strong passwords.

In our usability study, we asked users to use the encrypted link option with password protection in Tresorit and send the link by e-mail.

\subsection{Hypotheses}
Our expectations before the study were that users will find TwinCloud easier to use since it does not require to use an e-mail service to share a file. Moreover, we thought that the time spent to share a file will be less for TwinCloud. We also expected that users will find the Tresorit more secure as it has an additional password protection (One disadvantage of TwinCloud is that since its provided security is not directly observed by the users, they may find other methods with explicit security features more secure).
We specify our hypotheses for this study as follows:
\begin{itemize}
    \item [H1] Users would find TwinCloud more usable than Tresorit.
    \item [H2] Sharing time of TwinCloud will be smaller than the sharing time of Tresorit.
    \item [H3] Users would find Tresorit more secure.
\end{itemize}

\subsection{Methodology}
For our study we contacted 20 people that have not used Tresorit and TwinCloud before. We provided them the computer that has TwinCloud and Tresorit installed. We gave each user a scenario and tasks. We used System Usability Scale (SUS) \cite{23} to evaluate the usability of the programs.

A session starts with demographic questions, and continues with a scenario. Then, we gave users the tasks and the subtasks. After each task we asked the ten questions from SUS. 

\begin{figure}[!h]
\centering
\includegraphics[width=2.63in]{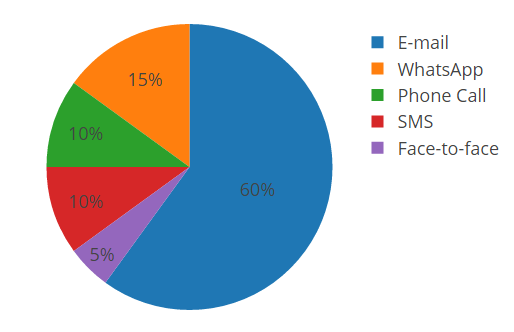}
\caption{Proportion of channels users choose to transmit passwords.}
\label{fig_9}
\end{figure}

Each task consists of two subtasks. First subtask asks users to share a file with a specific user and the second subtask asks users to open a file which is shared with them. For the tasks we gave them a start point such as the path of the file they need to share. Different from TwinCloud tasks, for Tresorit we asked users to specify the way they choose to transmit the password (Phone call, SMS, WhatsApp, E-mail, other). Fig. \ref{fig_9} demonstrate the proportion of channels users choose to transmit the passwords. (Majority of users preferred Email. When the password is shared through email, then combining encrypted link with the password protection offers no more security over encrypted link option alone.) At the end of the session we asked users to sort the systems by their usability in sharing and security.
\subsection{Results and analysis}
There were 20 users aged between 16 and 49 participated in this study. Among participants, 6 of them (30\%) were female. 11 users (55\%) were satisfied with their current skills for using Cloud Storage technologies.

\begin{figure}[!h]
\centering
\includegraphics[width=\columnwidth]{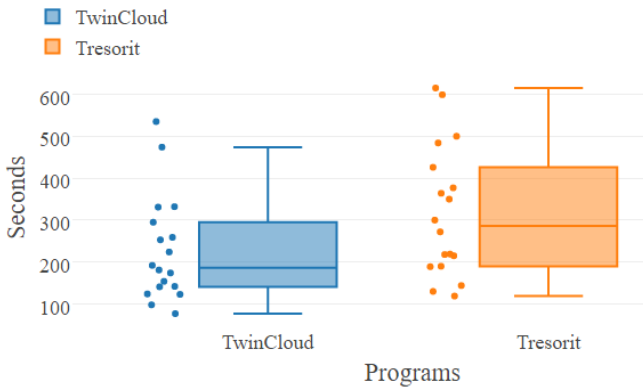}
\caption{Sharing times of TwinCloud and Tresorit.}
\label{fig_6}
\end{figure}

\begin{figure}[h]
\centering
\includegraphics[width=\columnwidth]{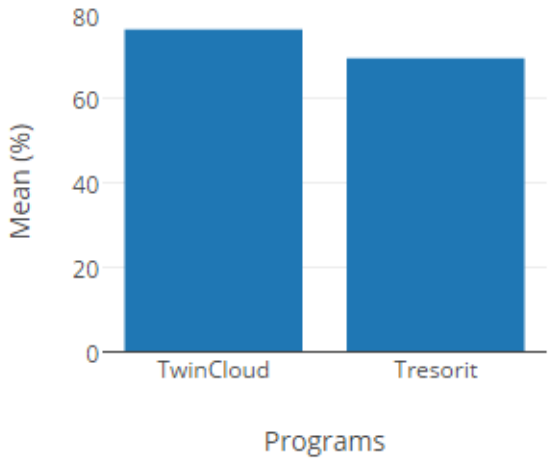}
\caption{Mean of SUS scores in percentage.}
\label{fig_7}
\end{figure}

\begin{figure}[h!]
\centering
\includegraphics[width=\columnwidth]{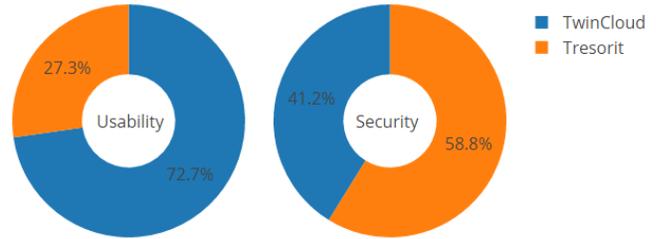}
\caption{Proportion of users who votes on TwinCloud and Tresorit in terms of usability and security.}
\label{fig_8}
\end{figure}

We first present the results in cumulative form for TwinCloud and Tresorit in Fig \ref{fig_6}, \ref{fig_7} and \ref{fig_8}. Fig. \ref{fig_6} shows the total time for sharing a file and opening a file shared with users. Fig. \ref{fig_7} compares the SUS scores of TwinCloud and Tresorit.

At the end of the session we asked users to choose the one that is more usable in sharing and the one that is more secure. Fig. \ref{fig_8} demonstrates the results. 72.7\% of users found TwinCloud more usable in sharing and 58\% of users found Tresorit more secure.

We used non-parametric tests in our analysis; the Wilcoxon signed-rank test for time comparison (H2), t-test for usability comparison (H1). A p-value less than 0.05 is considered as an evidence to reject the null-hypothesis.

According to Wilcoxon signed-rant test, the difference between sharing time of TwinCloud and Tresorit is significant (p = 0.022). This supports H2. 

The average SUS score of TwinCloud is 77.20\% and for Tresorit it is 68.67\%. Also, 72.7\% of users found TwinCloud more usable. This could be a supporting argument for our hypothesis (H1). However, according to the T-Test, the difference between the SUS scores of TwinCloud and Tresorit is marginally significant (p = 0.056). We think that this is because TwinCloud user interface does not look as professional as Tresorit. Also, during some of the sessions TwinCloud is slower because of slow internet connection. These kinds of disturbances might affect our results.

In our methodology users access files that are shared with them right after we send the encrypted link, so they easily find the link in their inbox. However in real situations users more probably spend more time searching the encrypted link in their inboxes. This may decrease the SUS scores of Tresorit and lengthen the time spent to open the shared files.

According to Fig. 8, More than half of the people (58.8\%) found Tresorit more secure because of the additional password protection mechanism. This supports H3. However, 12 users (60\% of total users) used the same e-mail to transmit the password which we think weakens the security dramatically. Also, nearly half of the users (45\%) chose weak passwords such as 123.

As a result of our usability study, we can conclude that our hypotheses are supported (H1 is only weakly supported).

\section{Extensions And Future Works}
To hide the real file names from cloud providers, we encrypt the file names while uploading to Google Drive and Dropbox. To do this, TwinCloud generates two encryption keys. We store the first one in Google Drive and use to encrypt the file names on Dropbox. The second key is stored in Dropbox and used to encrypt the file names on Google Drive. Moreover, to protect the integrity of a file, TwinCloud computes a Message Authentication Code (MAC) of the original file and uploads the result to Dropbox. The key used to calculate the MAC is stored in Google Drive.

For future work, for information-theoretical level of security instead of AES, One-time pads can be used to encrypt the files. We note that long key sizes are less of a problem and they can be easily stored in the cloud and shared with other users using our solution. Furthermore, our solution can easily be adapted to more than two cloud providers. For instance, two of the three cloud providers can be used to store the encryption key and the other one can be used to store the encrypted file where both keys are needed to decrypt the file.

\section{Conclusion}
In this paper, we introduced TwinCloud to provide an easy-to-use cloud experience for cloud users who are also sensitive to the security of files they upload. We described a novel approach that uses two or more cloud service providers to securely store and share files without complex key management operations. TwinCloud provides security by using one cloud provider to store the encrypted file and the other one to store the encryption key. It executes all the necessary file operation in the back-end and does not ask cloud users for complex key distribution operations. Since, TwinCloud creates all the cryptographic files on-the-fly and does not need to store encryption keys, encrypted files and user information, it is a lightweight solution. TwinCloud's usability and perceived security were evaluated and compared with Tresorit by a usability study. We concluded that TwinCloud is more usable for sharing, however, users think that it is less secure than Tresorit's password-protected encrypted link share method. TwinCloud is an easily applicable security solution in scenarios where it is assumed that cloud providers do not collude with each other.


\begin{thebibliography}{1}
\footnotesize
\bibitem{1}
R. Walters, "Cyber attacks on U.S. companies in 2014," Issue Brief No. 4289, October 2014. \\[-15pt]
\bibitem{2}
A. Ferdowsi, "Yesterday's Authentication Bug, Dropbox Blog," June 2011, Available: https://blogs.dropbox.com/dropbox/2011/06/yesterdays-authentication-bug/ (last access June 21, 2016).\\[-15pt]
\bibitem{3}
C. Louie, "Have you enabled two-step verification?," Dropbox Blog, October 2014, Available: https://blogs.dropbox.com/dropbox/2014/10/have-you-enabled-two-step-verification/ (last access June 21, 2016).\\[-15pt]
\bibitem{4}
Dropbox, "How secure is Dropbox?," Dropbox Help Center, Available: https://www.dropbox.com/en/help/27 (last access June 21, 2016).\\[-15pt]
\bibitem{5}
Google, "Google terms of service," April 2014, Available: https://www.google.com/intl/en/policies/terms/ (last access June 21, 2016).\\[-15pt]
\bibitem{6}
D. Sheng, D. Kondo, and F. Cappello, "Characterizing cloud applications on a Google data center," IEEE 42nd International Conference on Parallel Processing. Lyon, pp. 468-473, October 2013.\\[-15pt]
\bibitem{7}
Google, Server-Side Encryption, Google Cloud Platform, Available: https://cloud.google.com/datastore/docs/concepts/encryption-at-rest (last access June 21, 2016).\\[-15pt]
\bibitem{8}
J. Kincaid, "Google confirms that it fired engineer for breaking internal privacy policies," September 2010, Available: http://techcrunch.com/2010/09/14/google-engineer-spying-fired/ Chu (last access June 21, 2016).\\[-15pt]
\bibitem{9}
C. Kang Chu, W. Tao Zhu, J. Han, J. Liu, J. Xu, and J. Zhou, "Security concerns in popular cloud storage services," IEEE Pervasive Computing, October 2014.\\[-15pt]
\bibitem{10}
E. Duarte, F. Pinheiro, A. Zúquete, and H. Gomes, "Secure and trustworthy file sharing over cloud storage using eID tokens," OID conference, 2014.\\[-15pt]
\bibitem{11}
X. Chun Yin, Z. Guang Liu, and H. Jae Lee, "An Efficient and secured data storage scheme in cloud computing using ECC-based PKI," Advanced Communication Technology (ICACT) IEEE, 2014.\\[-15pt]
\bibitem{12}
G. Zhao, C. Rong, J. Li, F. Zhang, and Y. Tang, "Trusted data sharing over untrusted cloud storage providers," Cloud Computing Technology and Science (CloudCom) IEEE, 2010.\\[-15pt]
\bibitem{13}
Boxcryptor, Available: https://www.boxcryptor.com/en (last access June 21, 2016).\\[-15pt]
\bibitem{14}
Boxcryptor, Technical Overview, Available: https://www.boxcryptor.com/en/technical-overview (last access June 21, 2016).\\[-15pt]
\bibitem{15}
Tresorit, Available: https://tresorit.com/ (last access June 21, 2016).\\[-15pt]
\bibitem{16}
D. Wilson, and G. Ateniese, "To share or not to share in client-side encrypted clouds," Information Security Lecture Notes in Computer Science Volume 8783, 2014, pp. 401-41.\\[-15pt]
\bibitem{17}
[RFC6749] D. Hardt, "The OAuth 2.0 authorization framework", RFC 6749, October 2012.\\[-15pt]
\bibitem{18}
A. Holmes, and M. Kellogg, "Automating functional tests using Selenium," AGILE '06 Proceedings of the conference on AGILE, 2006.\\[-15pt]
\bibitem{19}
E. Macaskill, and G. Dance, "NSA Files: Decoded," November 2013, Available: http://www.theguardian.com/world/interactive/2013/nov/01/snowden-nsa-files-surveillance-revelations-decoded (last access June 21, 2016).\\[-15pt]
\bibitem{20}
A. Jivanyan, R. Yeghiazaryan, A. Darbinyan, and A. Manukyan, "Secure collaboration in public cloud storages," Volume 9334 of the series Lecture Notes in Computer Science, pp. 190-197, September 2015.\\[-15pt]
\bibitem{21}
P. Gharjale, and P. Mohod, "Efficient public key cryptosystem for scalable data sharing in cloud storage," Computation of Power, Energy Information and Communication (ICCPEIC), April 2015.\\[-15pt]
\bibitem{22}
C. Kang Chu, S. Chow, W. Guey Tzeng, J. Zhou, and R. Deng, "Key-aggregate cryptosystem for scalable data sharing in cloud storage", IEEE Transactions on Parallel and Distributed Systems Volume 25, pp. 468-477, April 2013.\\[-15pt]
\bibitem{23}
J. Brooke, "SUS – A quick and dirty usability scale", In P. W. Jordan, B. Thomas, B. A. Weerdmeester, and A. L. McClelland. Usability Evaluation in Industry. London: Taylor and Francis, 1996.\\[-15pt]
\bibitem{24}
D. Deniz Yavuz and S. Gurkan, TwinCloud, 2016, GitHub repository, https://github.com/DenizYavuz/TwinCloud\\[-15pt]
\bibitem{25}
Tresorit, Encrypted link Whitepaper, Available: https://tresorit.com/files/encrypted-link-whitepaper.pdf (last access June 14, 2016).\\[-15pt]
\bibitem{26}
Whitten, Alma, and J. Doug Tygar. Why Johnny Can't Encrypt: A Usability Evaluation of PGP 5.0. Usenix Security. 1999.\\[-15pt]
\end{thebibliography}
\end{document}